\documentclass[11pt]{article}
\usepackage{amsmath,amssymb,mathtools,amsfonts,epsfig,graphicx,euscript,psfrag,amsfonts,dcolumn,bm}
\usepackage{amsmath,amssymb}
\usepackage[colorlinks,linkcolor=blue,citecolor=magenta]{hyperref}
\newcommand{\bse}{\begin{subequations}}
\newcommand{\ese}{\end{subequations}}
\newcommand{\be}{\begin{equation}}
\newcommand{\ee}{\end{equation}}
\newcommand{\bea}{\begin{eqnarray}}
\newcommand{\eea}{\end{eqnarray}}
\newcommand{\ba}{\begin{array}}
\newcommand{\ea}{\end{array}}

\begin{document}

\thispagestyle{empty}

\vspace{2cm}
\begin{center}
\font\titlerm=cmr10 scaled\magstep4 \font\titlei=cmmi10
scaled\magstep4 \font\titleis=cmmi7 scaled\magstep4 {
\Large{\textbf{Spiraling  String in Gauss$-$Bonnet Geometry}
\\}}
\vspace{1.5cm}
 \noindent{{Mahdi Atashi, Kazem Bitaghsir Fadafan
}}\\
\vspace{0.8cm}
{\em
Faculty of Physics, Shahrood University of Technology, P.O.Box 3619995161,  Shahrood, Iran}\\

\vspace*{.25cm}
\textit{E-mail:} \verb|m.atashi@shahroodut.ac.ir|, \verb|bitaghsir@shahroodut.ac.ir|  
\vspace*{.25cm}

\vspace*{.4cm}

\end{center}
\vskip 2em

\begin{abstract}
In this paper, we consider a spiraling string falling in the bulk with Gauss$-$Bonnet geometry that is holographically dual to a heavy particle rotating through a hot plasma at finite coupling.  One finds such interesting simple problem provides a novel perspective on different channels of the energy loss in  the corresponding strongly coupled theory.  Depends on the sign of the coupling, one observes that the influence of finite coupling on total energy loss and contribution of drag force and radiation channels appears as a shift on curves with respect to the plasma with infinite coupling. Also we found that crossover between regime in which drag force contribution is predominant to regime in which energy loss is due to radiation, does not depend on the Gauss$-$Bonnet coupling.  

\end{abstract}
\section{Introduction}
The  AdS/CFT correspondence is a very  useful technique to describe a strongly-coupled gauge theory by a classical theory of gravity in a  higher dimensional space time \cite{CasalderreySolana:2011us}. It has been used to derive universal properties of these theories like the ratio of the shear viscosity to the entropy density of the hot plasma \cite{DeWolfe:2013cua}. Using this correspondence, we study the energy loss of an accelerated heavy point particle in the Gauss-Bonnet background. This particle is the end point of a semi-classical string  spiraling in the bulk geometry. We fix the Guass-Bonnet coupling constant, $\lambda_{GB}$ and assume that the probe particle rotates along a circle of radius $l$ with an angular frequency $\omega$. The constant velocity and acceleration of this particle are given as $v=l\omega$ and $a=l\omega ^2$, respectively. \\

Two different channels of the energy loss are drag force and radiation. Using the numeric, we verify that the rate of the energy loss is similar to the drag force formula at constant velocity and small acceleration where $a=v\omega\rightarrow 0$, meaning $\omega\rightarrow 0$. The drag force formula in Gauss-Bonnet geometry has been studied in \cite{Fadafan:2008gb,VazquezPoritz:2008nw}. One expects the other channel of energy loss i.e., radiation by increasing acceleration of the particle. One reaches this limit either by decreasing $l$ ($l\rightarrow 0$) or increasing $\omega$ ($\omega \rightarrow \infty$). However, based on our knowledge there is no analytic formula for radiation of an accelerated particle in Gauss-Bonnet background \footnote{ The general result for radiation of an accelerated particle in $\mathcal{N}=4$ SYM vacuum has been derived in \cite{Mikhailov}. }. One should notice that we cannot separate out two different channels sharply as it was discussed in \cite{Fadafan:2008bq}. But it is still useful to understand the effect of $\lambda _{GB}$ of the background. The Gauss-Bonnet corrections include curvature squared terms provide a good example of higher derivative corrections where in the resulting action there is no ghost. One important observation in this case is violation of the bound on the shear viscosity to entropy density, $\frac{\eta}{s}$ in CFTs dual to Gauss-Bonnet gravity \cite{Brigante:2007nu,Brigante:2008gz}. Although, the theory may be inconsistent regarding micro-causality \cite{Camanho:2009vw}.  Based on the holography such corrections correspond to large 't Hoof coupling $\lambda$ in the boundary field theory. The study of finite coupling corrections on the fast thermalization process has been done in \cite{self2,Grozdanov:2016zjj}. The effect of higher derivative  corrections on the different aspects of heavy quarks in the Quark-Gluon Plasma(QGP) has been studied in\cite{Fadafan:2008gb,Fadafan:2013coa,Fadafan:2015kma,Fadafan:2012qy,Fadafan:2011gm,AliAkbari:2009pf,Fadafan:2008uv}. Authors in \cite{McInnes:2018ibt} considered the effect of vorticity on the thermodynamics of the QGP. Also, the study of holographic multi-quark states in the deconfined QGP has been done in 
\cite{Burikham:2009yi,Burikham:2010sw}.\\

The problem of a spiraling string in different bulk geometries reveal the importance of this simple set up for studying the energy loss in the corresponding field theories. For example, this problem in Schrodinger or Lifshitz  field theories have been studied in \cite{Akhavan:2008ep,Fadafan:2009an,Hartnoll:2009ns,Alishahiha:2012cm,self1,Kiritsis:2012ta}. Interestingly,  we found that at zero temperature there is a critical radius $l_c$ where the total energy loss of the rotating particle does not depend on the non-relativistic parameters of the theory. For spiraling string in confining, anisotropic and non-conformal geometries see \cite{AliAkbari:2011ue,Fadafan:2012qu,Ewerz:2012fca} respectively. One finds such interesting simple problem provides a novel perspective on different channels of the energy loss in strongly coupled field theories. \\

The study of energy loss is an interesting and important problem in studying quark-gluon-plasma (QGP) produced at RHIC and LHC \cite{Matsui:1986dk}. In this case the point particle could be a heavy quark. Study of such problems need non perturbative strongly coupled approaches and time dependent methods, then using the AdS/CFT correspondence is reliable \cite{CasalderreySolana:2011us,DeWolfe:2013cua}. First study of the energy loss of heavy quarks from the drag force channel has been done in \cite{Herzog:2006gh,Gubser:2006bz}. \\

Here, we do not consider the back-reaction of the heavy particle on the boundary theory. This approximation leads to absence of the broadening in the angular distribution of radiated power in strongly coupled field theory   \cite{Athanasiou:2010pv,Chesler:2011nc}. It was argued that the super-gravity approximation is responsible for this phenomena  \cite{Hatta:2010dz,Agon:2014rda}. Study of the energy loss from accelerating objects is basic and interesting problem of quantum field theories. It is very difficult to describe it in the strongly coupled systems. Finding a framework to describe radiation in the Gauss$-$Bonnet background would be very interesting.\\

This paper is organized as follows. In section two, we will
present the gravitational dual of a rotating heavy particle at finite coupling, and discuss about worldsheet horizon behavior. Next, we study drag force and radiation contributions in total energy lost by the rotating particle for different values of the Gauss$-$Bonnet coupling $\lambda _{GB}$. In the conclusion section, we discuss about crossover between drag and radiation channels and regime of validity of semi$-$classical calculation at finite coupling.\\
\section{Holographic Setup}

Now we shall consider a heavy rotating particle at finite coupling and  finite temperature. Holographically, considering finite coupling field theory is dual to add higher derivative gravity terms in gravitational geometry and finite temperature is dual to creating black hole in the bulk. Here we concentrate on Gauss$-$Bonnet AdS-Schwarzschild black hole as gravitational dual with the metric \cite{Cai:2001dz}\\

\be \label{metric at T}
ds^2=\frac{1}{u^2\,L^2}\left(-f(u)\,dt^2+\frac{du^2}{f(u)}+\sum _{i=1}^3\,dx_i^2\right).\,\,\,\,\,\,
\ee \\

where\footnote{The factor $L$ is related to the AdS curvature. We set $L=1$ in this paper.}\\

\be
f(u)=\frac{n}{2 \lambda _{GB}}\left( 1-\sqrt{1-4 \lambda _{GB} \left(1-(\frac{u}{u_h})^4\right)}\right),
\ee \\

that is normalized by $n=\frac{1}{2}(1+\sqrt{1-4 \lambda _{GB}})$  that made speed of light to unity on the boundary \cite{Buchel:2009}. Also $\big( u,  x_i  \big) $ are inverted bulk radial and boundary spatial  coordinates respectively. The boundary is located at $u = 0$ and horizon is denoted as $u_h$ which can be  found by solving $f(u_h)=0$. \\

The temperature is given by\\

\be
T=\frac{1}{\pi\, u_h \sqrt{n}}.
\ee \\

We fix temperature as $T=\frac{1}{\pi}$ and the horizon position $u_h$ depends on $\lambda _{GB}$ as $u_h=\frac{1}{\sqrt{n}}$. Also the mass of the black hole is $M=\frac{3 \Sigma _k}{16\pi\,G\,u_h^4}$, where $\Sigma _k$ is volume of co-dimension two  hyper-surface  at the boundary \cite{Cai:2001dz}.  \\

First, we should use the ansatz $X^{\mu}\equiv \left( t=\tau ,u=\sigma ,\rho=\rho (u),\phi=\omega \tau +\phi (u),x_3=0\right)$, where $(\rho , \phi)$ are radial and angular profiles of the spiraling string respectively, so that $\rho (0)=l$ and $\phi(0)=0$, and $ \mu $ runs over space-time coordinates.\\

Also the  Nambu-Goto action\\

\be
S_{\text{NG}} =-\frac{1}{2\pi\,\alpha'} \int d\tau\,d\sigma \sqrt{-g},
\ee \\

 is used to study the shape of the spiraling string in the bulk, where $g$ is determinant of induced metric $g_{\alpha \beta}=\left( \partial _{\alpha}X^{\mu}\partial _{\beta}X^{\nu}\right) G_{\mu \nu}$. The $ G_{\mu \nu} $ is the metric \eqref{metric at T} in cylindrical coordinate. It is easy to compute the elements of induced metric those are \\
 
 \be \label{wshmetric}
 \begin{array}{cc}
g_{\tau \tau}=\frac{-f(u)}{u^{2}}+\frac{\rho^2\omega^2}{u^2}, & g_{uu}=\frac{1}{u^2 f(u)}+\frac{\rho^2}{u^2}\phi '^2+\frac{\rho '^2}{u^2},\\
 g_{\tau u}=g_{u \tau}=\omega ^2\,\frac{\rho^4}{u^4}\,\phi'^2, & g_{ab}=0,
 \end{array}
 \ee \\ 
 
where $a$ and $ b $ run over $ (\rho , \phi ,z) $. So determinant of the induced metric is $ g=g_{\tau \tau}\,g_{uu}-g_{\tau u}^2 $, and Lagrangian density, $ \mathcal{L}=\sqrt{-g} $, is given by\\

\be
\mathcal{L} = u^{-2}\sqrt{\left(f(u) - {\rho^2}{\omega^2}\right)\left(\frac{1}{f(u)} +
	{{\rho'}^2}\right) + {\rho^2}{{\phi'}^2}f(u)}.
\ee \\

We assume the constant of motion as $\Pi=-\frac{\partial \mathcal{L}}{\partial \phi^{\prime}} $, then the equation of motion of $\phi(u)$ is given by\\

\be \label{phiT}
\phi'^2=\Pi ^2\frac{\left(\frac{-f(u)}{u^{2}}+\frac{\rho^2\omega^2}{u^2}\right)\left(\frac{1}{u^2f(u)}
	+\frac{\rho'^2}{u^2}\right)}{\left(\frac{-\rho(u)^2f(u)}{u^{4}}\right)\left(-\rho(u)^2f(u)u^{-4}
	+\Pi^2\right)}.
\ee \\

The right hand side of \eqref{phiT} must be positive. So there is a special point in bulk hologram dimension, $u_t$, that one finds it by solving the following equations:\\

\be \label{realitycond}
\frac{-f(u_t)}{u_t^{2}}+\frac{\rho_t^2\omega^2}{u_t^2}=0,\,\,\,\,\,\,\,\,\,\,\,\,
-u_t^{-4}\rho _t^2f(u_t)+\Pi^2=0,
\ee \\

which means that region $u>u_t$ is not considerable physically.\\ 

Comparing \eqref{wshmetric} and \eqref{realitycond} , it is straightforward to find that $g_{\tau \tau}=0$ at $ u=u_t $. So let us to call $u_t$ as \textit{worldsheet horizon}. In other word, the $u_t$ disconnect region $u<u_t$ from the lower part of the string $u>u_t$ whose local velocity is higher than local velocity of light \cite{Fadafan:2008bq,Kundu:2018}.\\   

We can use \eqref{phiT} to  obtain the equation of motion for $\rho(u)$ in terms of the constant $\Pi $ as\\

\begin{align}\label{ODET}
&\rho''(u)+\frac{\rho(u)\left(1+f(u)\rho'(u)^2\right)\left(4f(u)\rho(u)\rho'(u)+u\left(
	2-\rho(u)f(u)\rho'(u)\right)\right)}{2u\left(-\Pi^2u^4+f(u)\rho(u)^2\right)}+\nonumber\\
&\frac{2+\rho(u)f'(u)\rho'(u)+2f(u)\rho'(u)^2+\omega^2\rho(u)^3f'(u)\rho'(u)^3}{2\rho(u)\left(f(u)-\rho(u)^2\omega^2\right)}=0.
\end{align}\\

The extracted equation of motion \eqref{ODET} is in agreement with the master equation found in \cite{self1} for $\theta=0,d=3$ and $z=1$.\\

One can use the special point $(\rho_t,u_t)$ as initial value for the differential equation of motion and solve it. The second initial condition $\rho'_t$ can be found from the prescription of \cite{Fadafan:2008bq} by using an expansion of $\rho(u)$ around $u=u_t$. One finds that the differential equation itself determines $\rho'_t$, one  obtains $\rho'_t$ by solving the following equation:\\

\be
-2\sqrt{f(u_t)}\omega+\left(-f(u_t)f'(u_t)+2u_t \,\omega^2\right)\rho'(u_t)+2 f(u_t)^{3/2}\,\omega\,\rho'(u_t)^2=0.
\ee
\\\\

\begin{figure}[ht]
	\includegraphics[width=16cm]{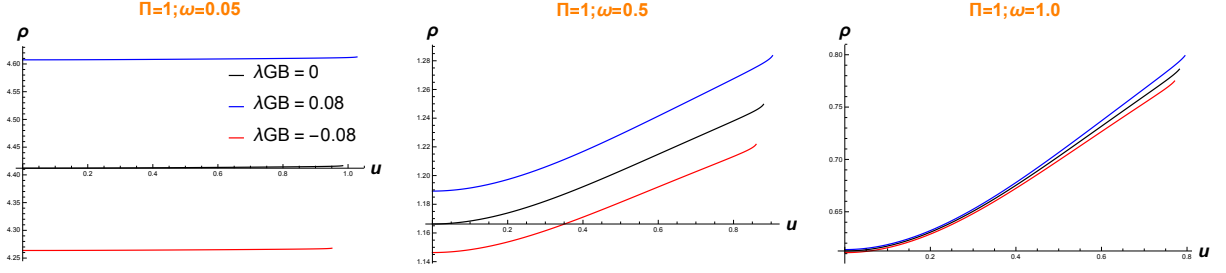}  
	
	\caption{\label{f1}  String radius vs radial direction in different $\lambda _{GB}$  with fixed $\Pi $. Plots from left to right are correspond to $\omega=0.05,0.5,1.0$ respectively.
		In each panel , curves from top to bottom  correspond to $\lambda _{GB}=0.08,0.0,-0.08$ respectively.}
\end{figure}

\textbf{\textit{String Shape:}} Having  values of $\rho(u_t)$ and $\rho'(u_t)$ as boundary condition, one can solve \eqref{ODET} numerically by \verb NDSolve   command of \verb Mathematica.    Fig. \ref{f1} shows the string shape in the bulk, $\rho (u)$ for different values of $\lambda _{GB}$. This figure is plotted for fixed $\Pi=1$ and panels from left to right are for  angular  velocity $\omega=0.05,0.5,1.0$, respectively.  In each panel, curves from top to bottom correspond to $\lambda _{GB}=0.08,0.00,-0.08$. The analysis of the graphs could be itemized as follow:     \\

\begin{itemize}
	\item At fixed $\Pi$ and $\omega$, the radius of rotation of the particle is greater (smaller) for positive (negative) $\lambda _{GB}$ than case correspond to $\lambda _{GB}=0$.
	
	\item At fixed $\Pi$, by increasing $\omega$,  curves correspond to different values of $\lambda _{GB}$ become closer together.
	
	\item At fixed $\omega$, by increasing $\Pi$, difference between curves correspond to different values of $\lambda _{GB}$ become larger.
	
	\item By increasing angular velocity $\omega$, curves associated with different $\lambda _{GB}$ become closer to each other. It means that gravitational geometry is $\lambda _{GB}-$independent at high $\omega$.
\end{itemize}

\begin{figure}[ht]
	
	\includegraphics[width=16cm]{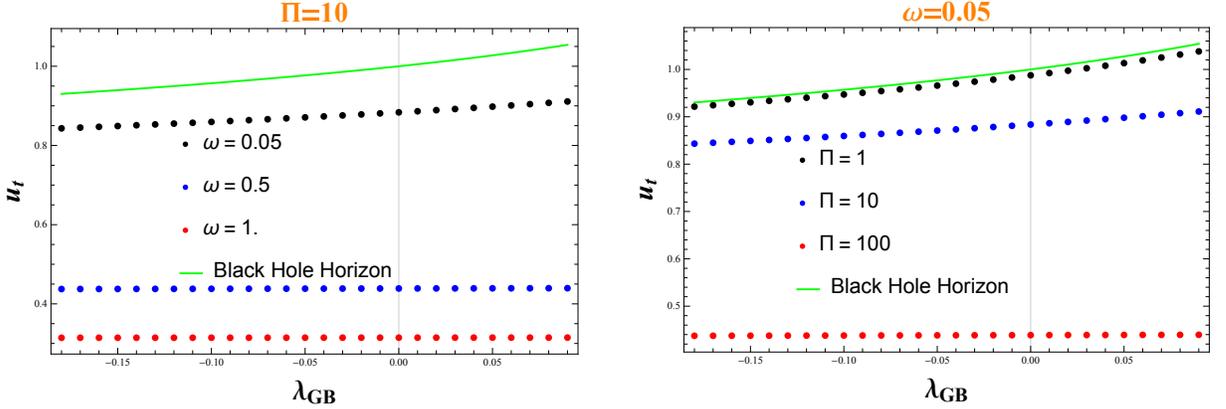} 
	
	\caption{\label{f2} Worldsheet horizon vs. $\lambda _{GB}$. The left panel shows behavior of worldsheet horizon vs. $\lambda _{GB}$ in different values of $\omega$, and the right one for different values of $\Pi$. At a constant $\lambda _{GB}$, for smaller value of $\omega , \Pi$ the worldsheet horizon located at deeper radial position.}
\end{figure}

\textbf{\textit{Worldsheet Horizon:}} Stochastic forces on a heavy particle lives on the boundary is related to the dual string worldsheet horizon \cite{Kundu:2018,Giataganas:2018rbq}. Therefore study of worldsheet behavior could be significant. The part of spiraling string that fall behind the worldsheet horizon does not have physical importance. Fig. \ref{f2} shows the radial position of the worldsheet horizon versus $\lambda _{GB}$ at different situations. In the left panel that is plotted for fixed $\Pi=10$, each curve from top to bottom correspond to $\omega =0.05,0.5,5.0$ respectively. As it is seen again, worldsheet horizon position does not depend on $\lambda _{GB}$ at high $\omega$. Also it is shown that, at low $\omega$, the worldsheet horizon will be located at deeper position in the bulk which is closer to black hole horizon. The right panel is similar to the left one, except that curves are classify according to different values of $\Pi=1,10,100$ from top to bottom and the figure is plotted in fixed angular velocity $\omega =0.05$. Similar to the left panel, the worldsheet horizon for smaller value of $\Pi$ becomes closer to the black hole horizon. Therefore we can summarize that the worldsheet horizon will be located at closer to the black hole horizon for smaller value of energy loss $\Pi \omega$.\\

Independence of the theory to $\lambda _{GB}$ at high $\omega$ could be seen in Fig. \ref{f2} too. In high $\omega$ and high $\Pi$ (and consequently at high $\Pi \omega$) the radial position of the world sheet horizon is constant over all the range of $\lambda _{GB}$.

\section{Energy Loss}
In this section, we study two different channels of energy loss, drag force and radiation. The total energy loss is given in terms of the the constant $\Pi$ as
\be
\frac{dE}{dt}=\frac{\pi}{2 }\sqrt{\lambda}\,T^2\,\, \Pi \, \omega ,
\ee
where  $\lambda$ is the 't Hooft coupling, and $T$ is temperature of the black hole. To find energy loss $\Pi \omega$ versus radius of rotation of heavy particle on the boundary, we choose different values of $\Pi$ and $\omega$ then by solving \eqref{ODET} numerically, one find $\rho (0)=l$ at each value for $\lambda _{GB}$. \\

\textbf{\textit{Drag force channel:}} Because of moving the particle at finite temperature  strongly coupled field theory, one expects that the particle experiences the linear drag force . Now we want to  study if the same mechanism exists for the total energy loss of the particle, i.e it looses the energy due to the linear drag force. The drag force on a moving heavy point particle in the Gauss-Bonnet gravity has been computed in \cite{Fadafan:2008gb}. The final result is given by

\be \label{dragcontribution}
(F_{drag})_{GB} = \frac{\pi}{2 }\sqrt{\lambda}\,T^2 \frac{ v }{\sqrt{n(n-v^2)+\lambda _{GB}v^4}},
\ee

where $v=l\omega$ is linear velocity of rotating particle. One finds the detail of drag force calculations in \cite{Fadafan:2009an}. \\

\textbf{\textit{Radiation channel:}} Since a heavy rotating quark is an accelerated charged particle, the radiation channel has contribution in total energy loss. The general Mikhailov's result for an accelerated particle in vacuum $\mathcal{N}=4$ SYM theory that is calculated in \cite{Mikhailov} is 

\be \label{rad}
\frac{dE}{dt}\vert _{vac.\, rad.}=\frac{\sqrt{\lambda}}{2 \pi}\frac{v^2 \omega ^2}{(1-v^2)^2}.
\ee

We calculate the ratio of enegy loss by radiation over total energy loss as 

\be\label{radiation}
\frac{\frac{dE}{dt}_{total}}{\frac{dE}{dt}_{rad}}=\frac{\Pi}{v^2 \omega}\left( 1-v^2 \right)^2.
\ee

\begin{figure}[ht]
	\centerline{\includegraphics[height=2.5cm,width=5.5cm]{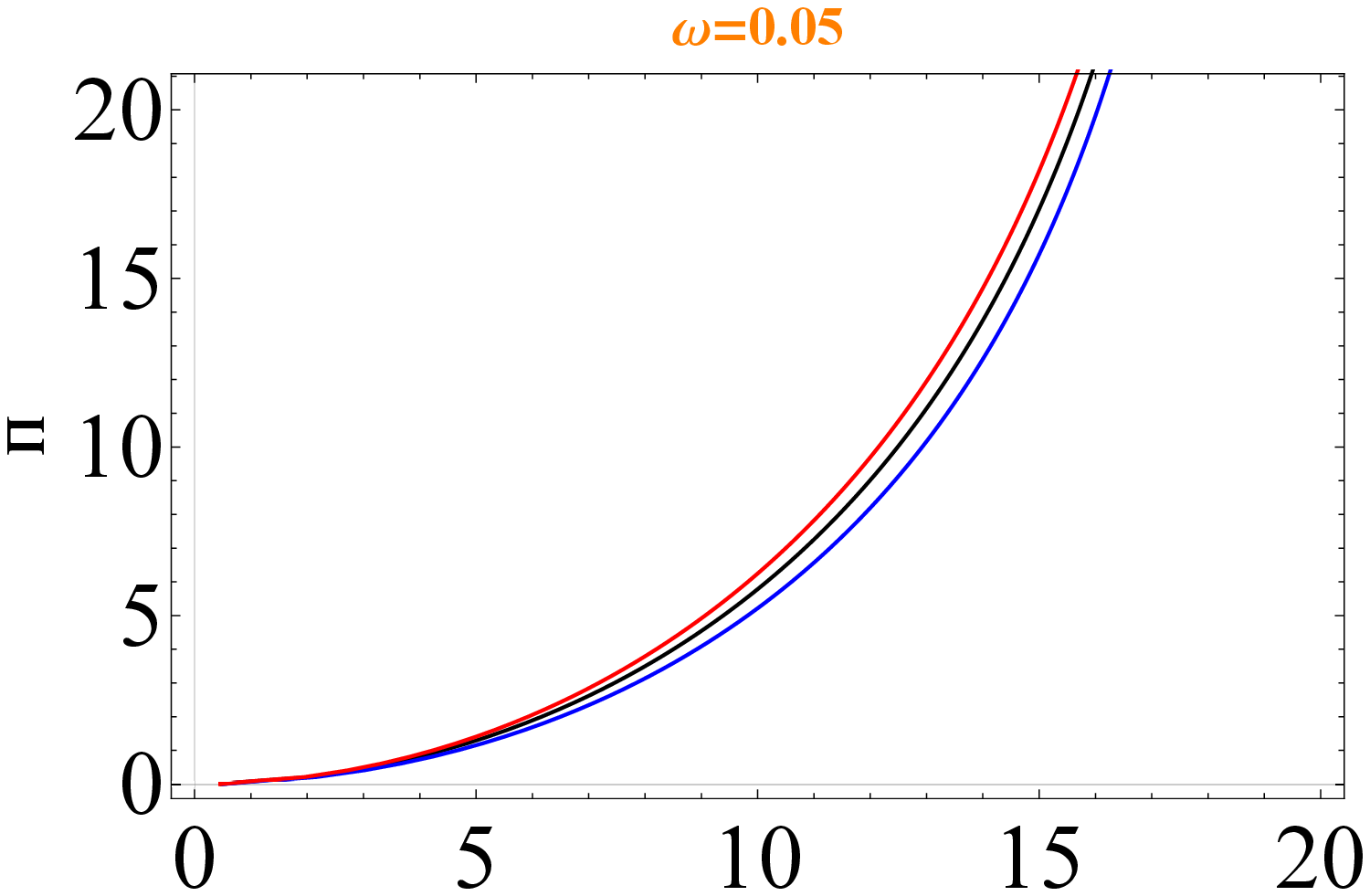}\includegraphics[height=2.5cm,width=5.5cm]{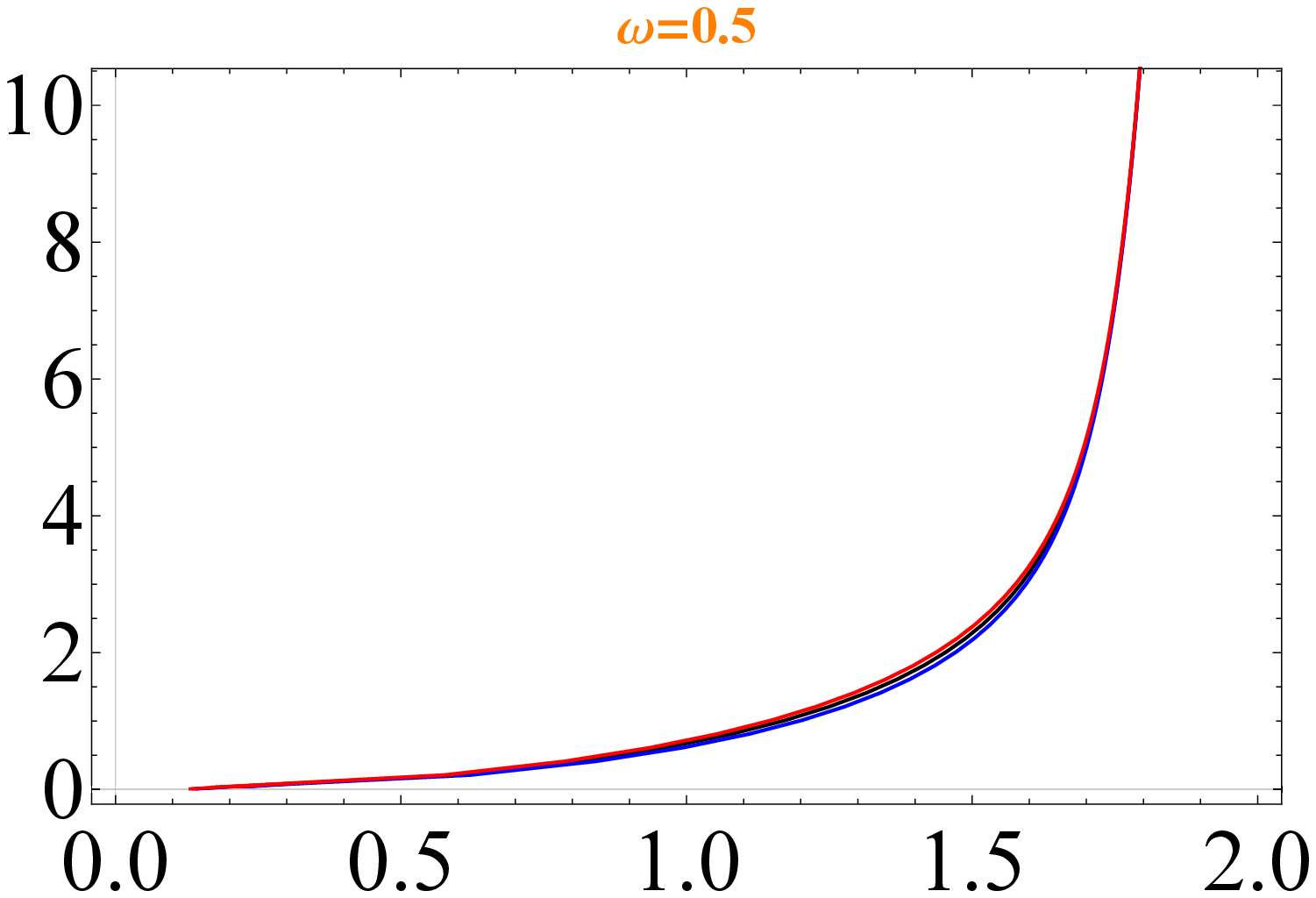}\includegraphics[height=2.5cm,width=5.5cm]{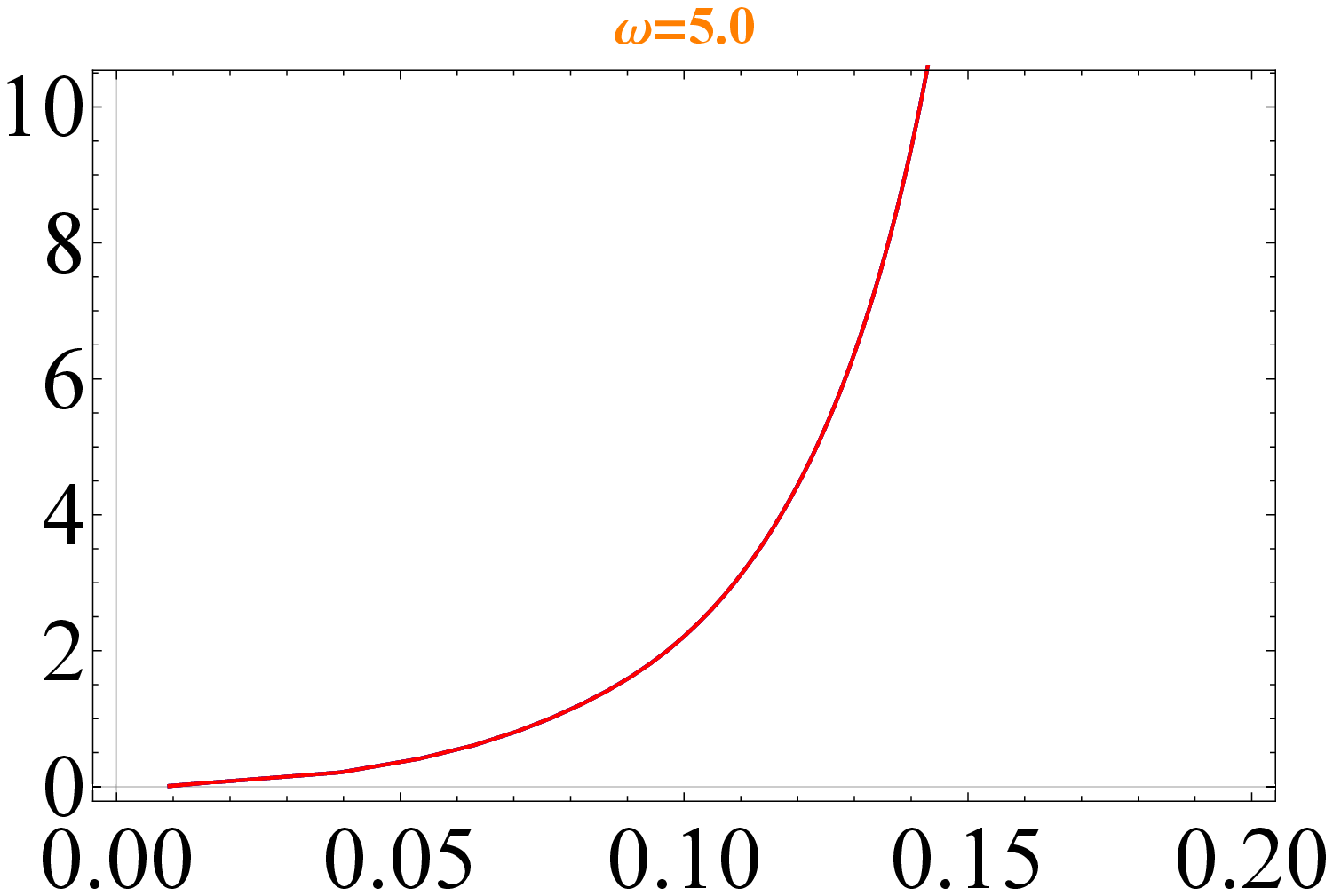} \\} 
	\centerline{\includegraphics[height=2.5cm,width=5.5cm]{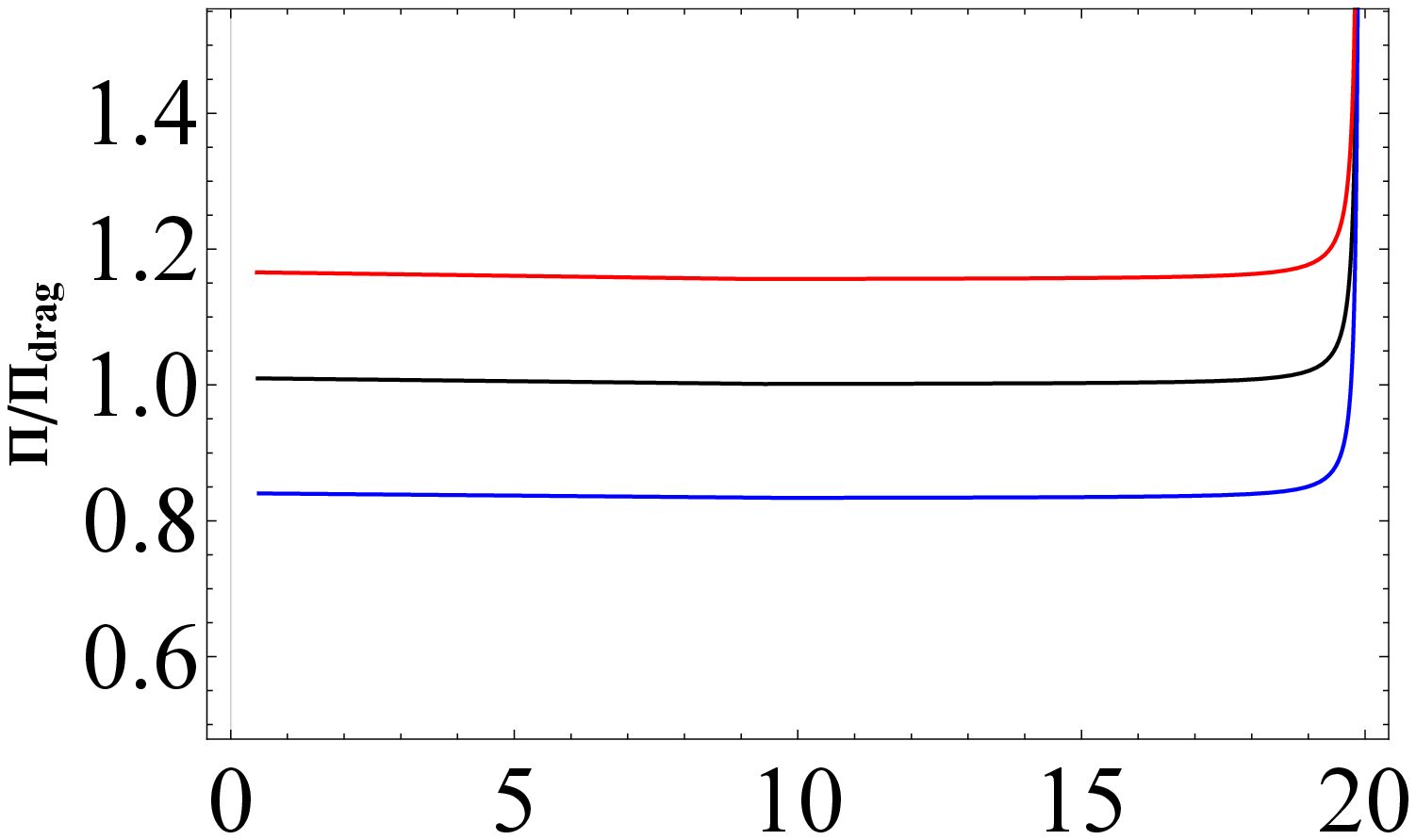}\includegraphics[height=2.5cm,width=5.5cm]{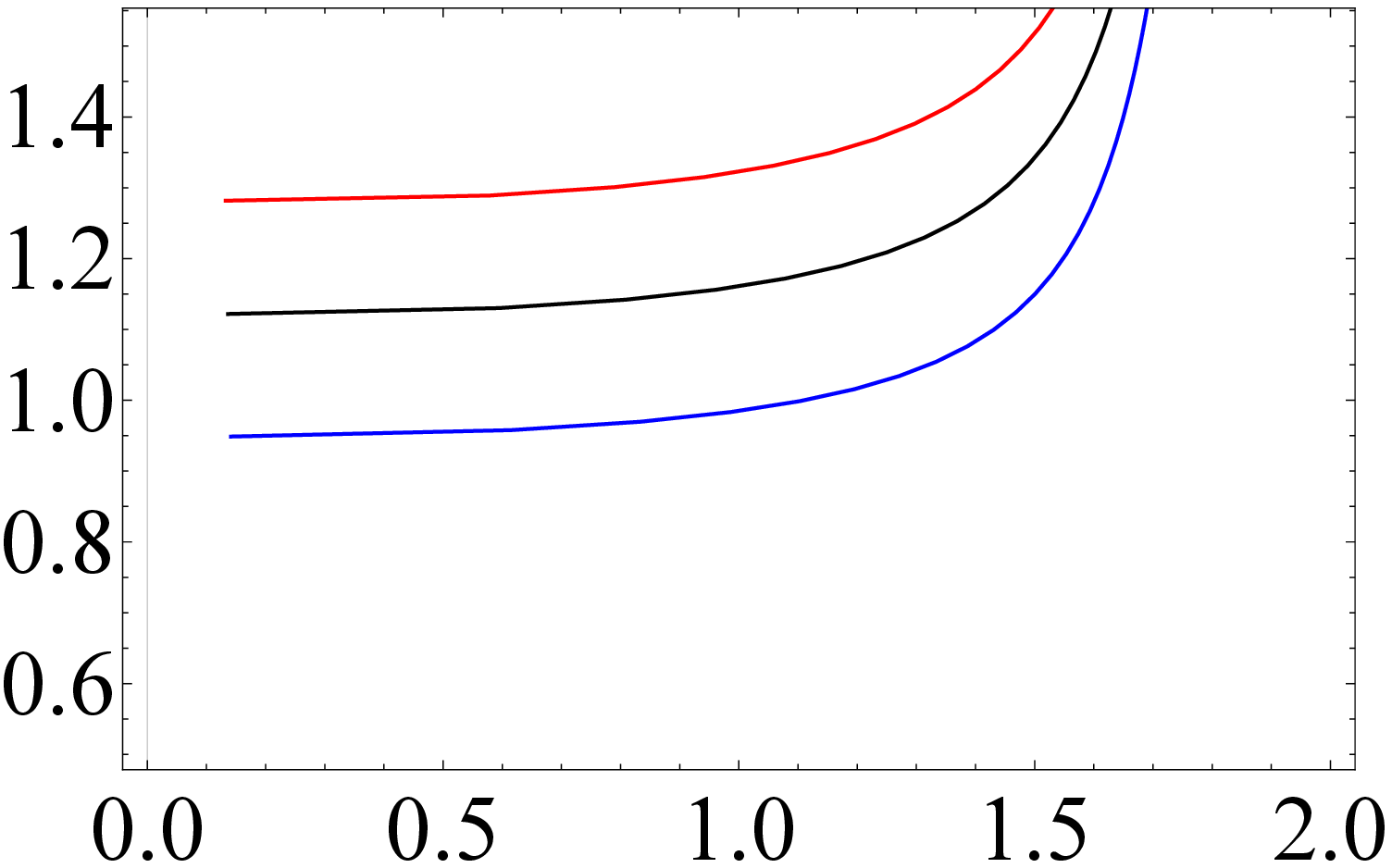}\includegraphics[height=2.5cm,width=5.5cm]{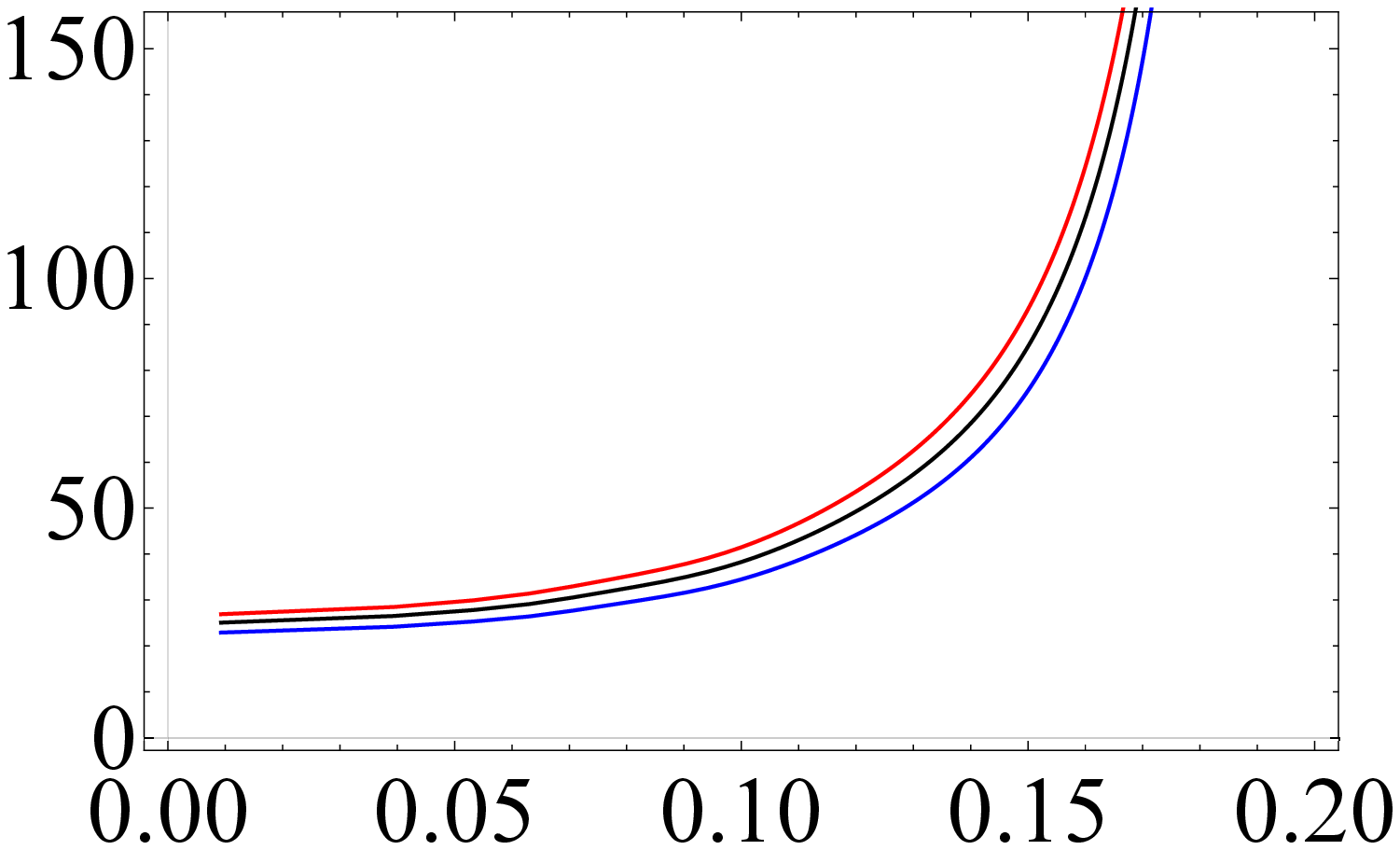} \\} 
	\centerline{\includegraphics[height=2.5cm,width=5.5cm]{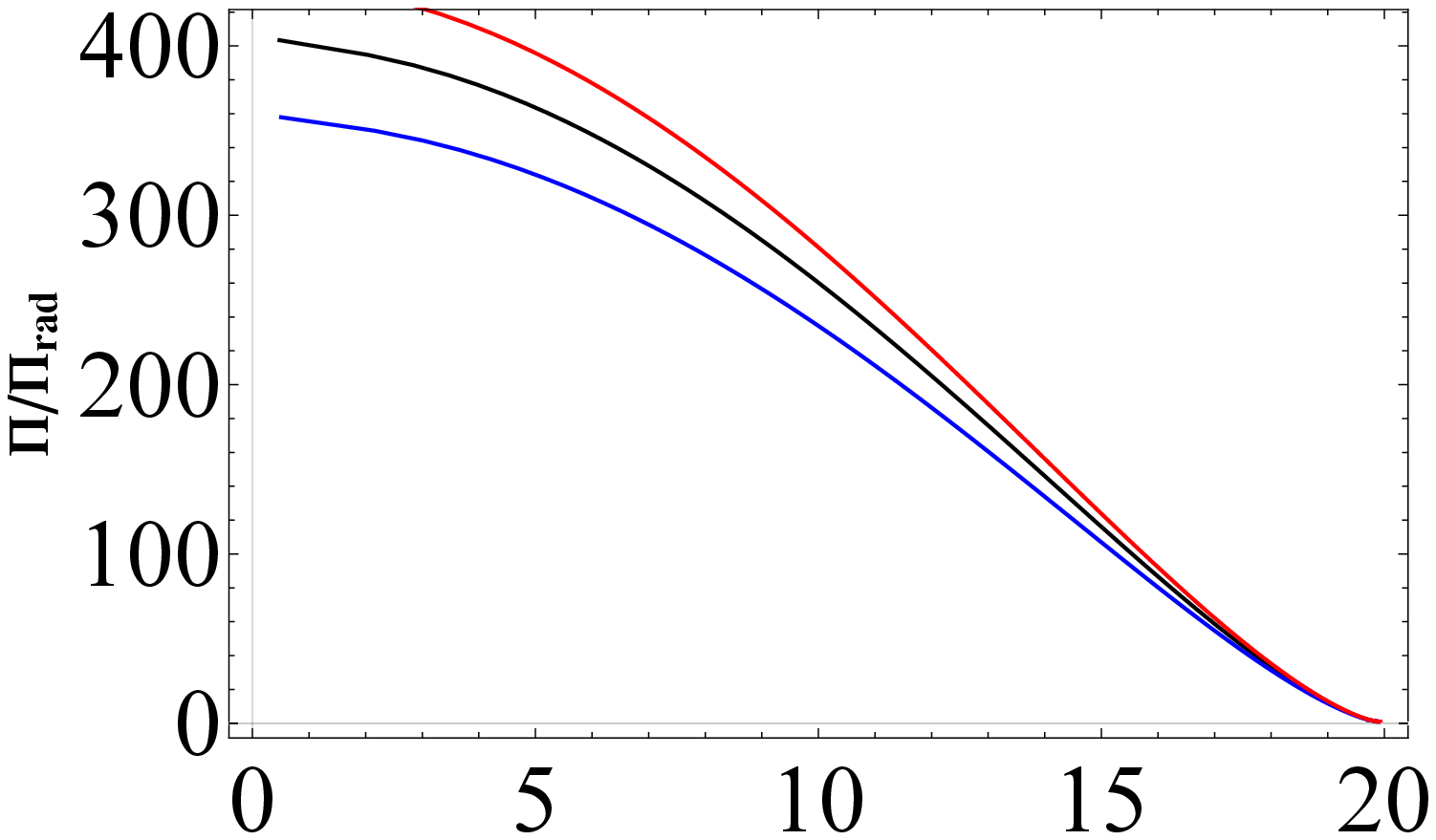}\includegraphics[height=2.5cm,width=5.5cm]{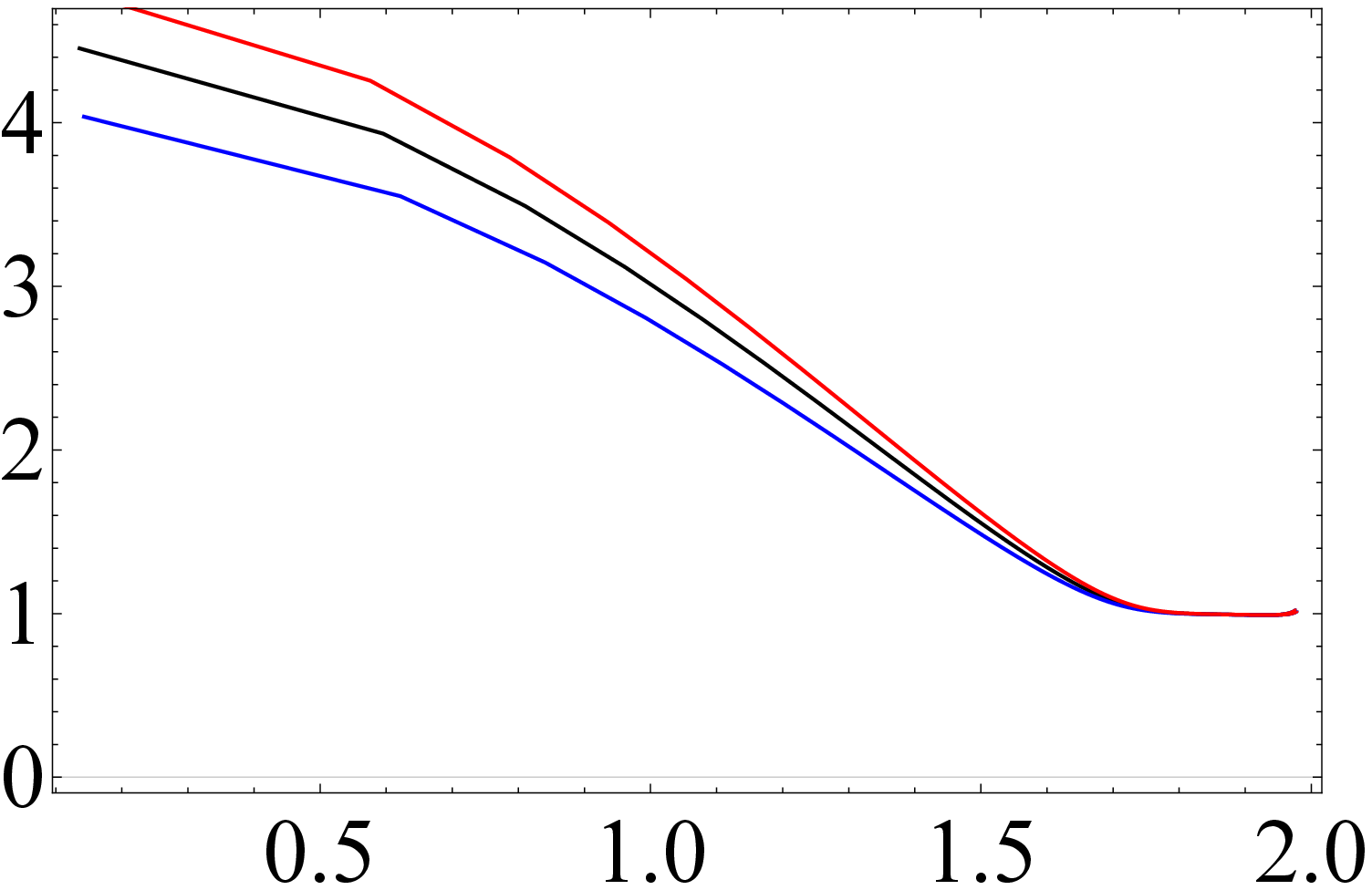}\includegraphics[height=2.5cm,width=5.5cm]{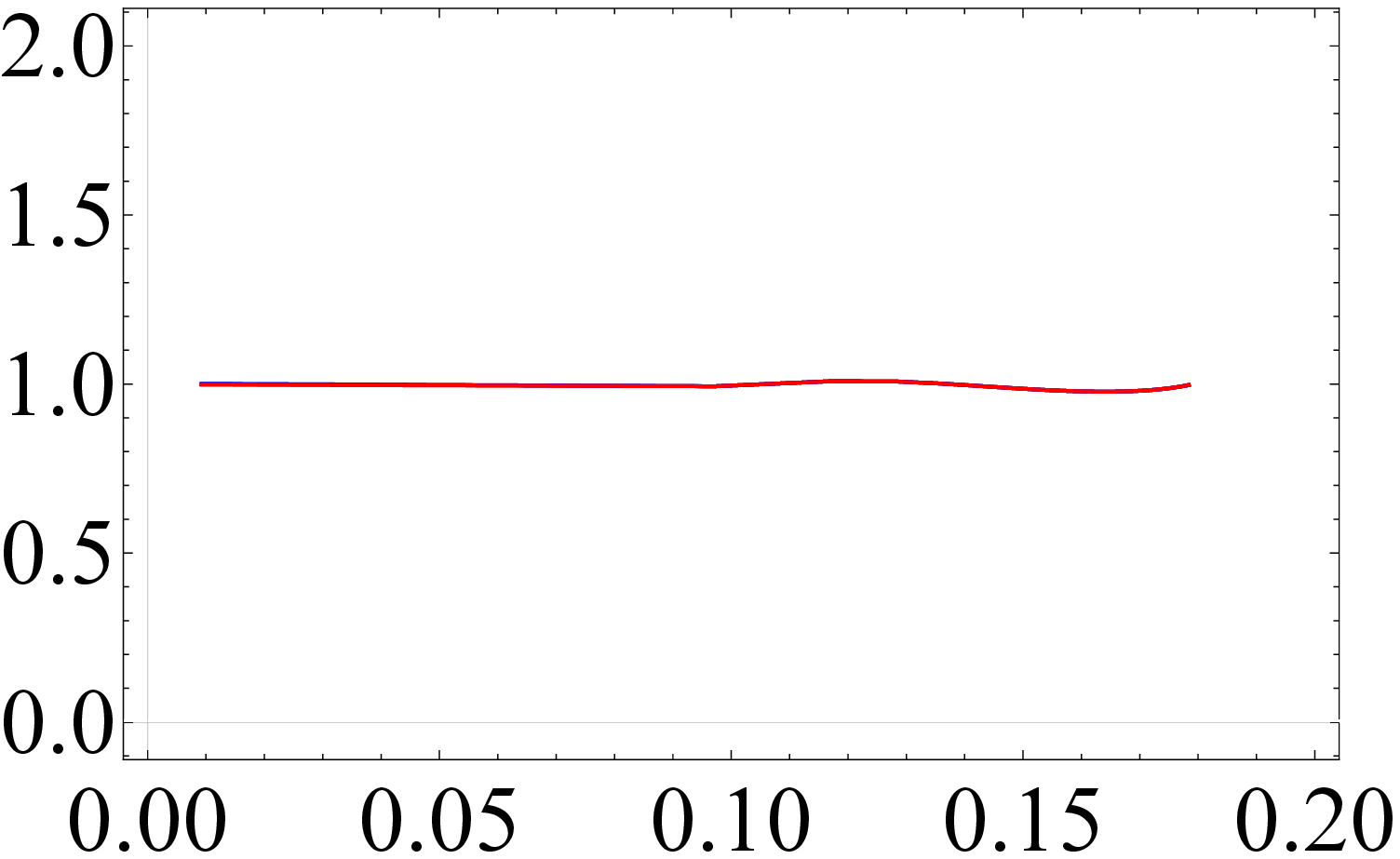} \\ }
	\centerline{\includegraphics[height=2.5cm,width=5.5cm]{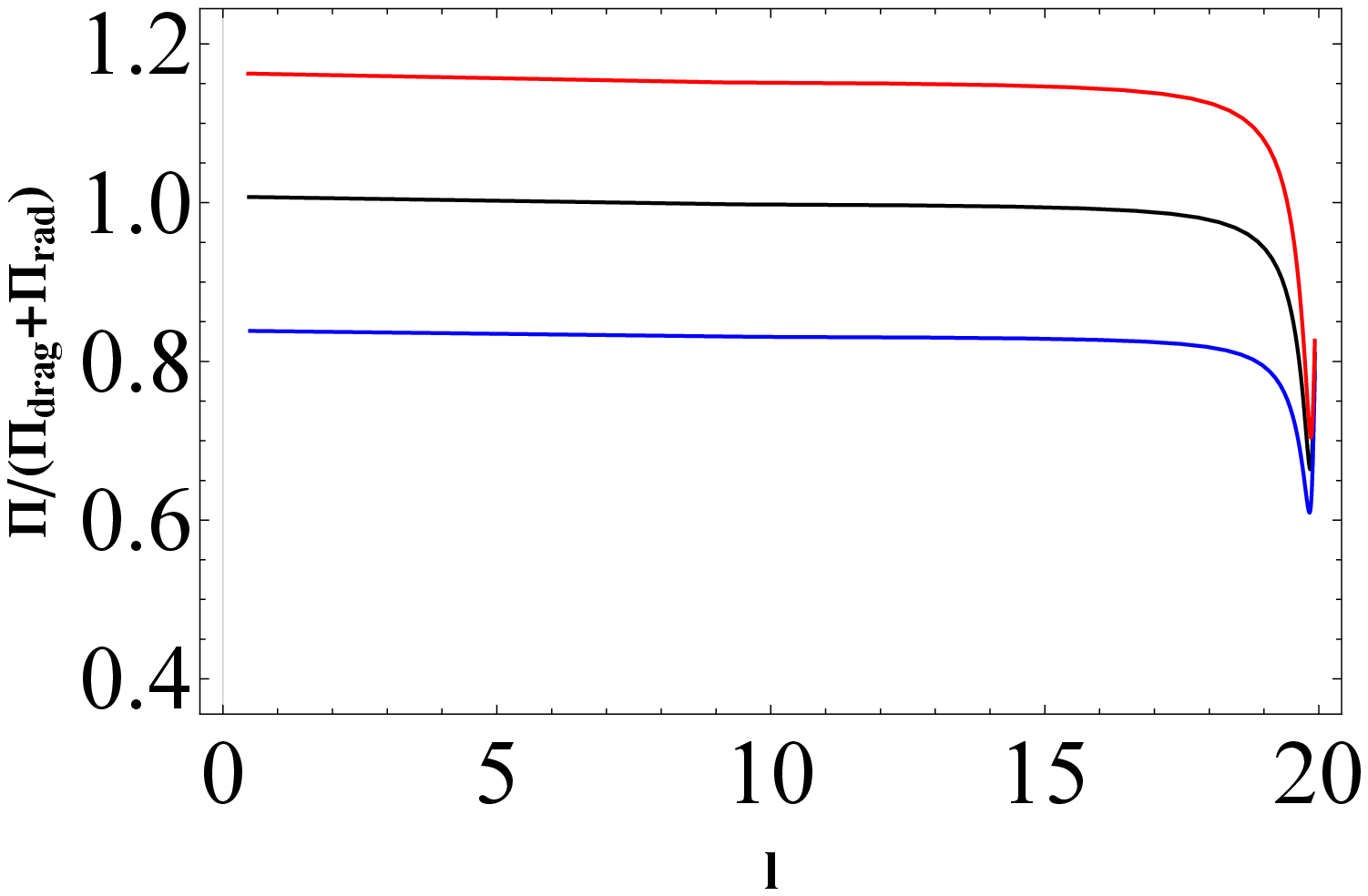}\includegraphics[height=2.5cm,width=5.5cm]{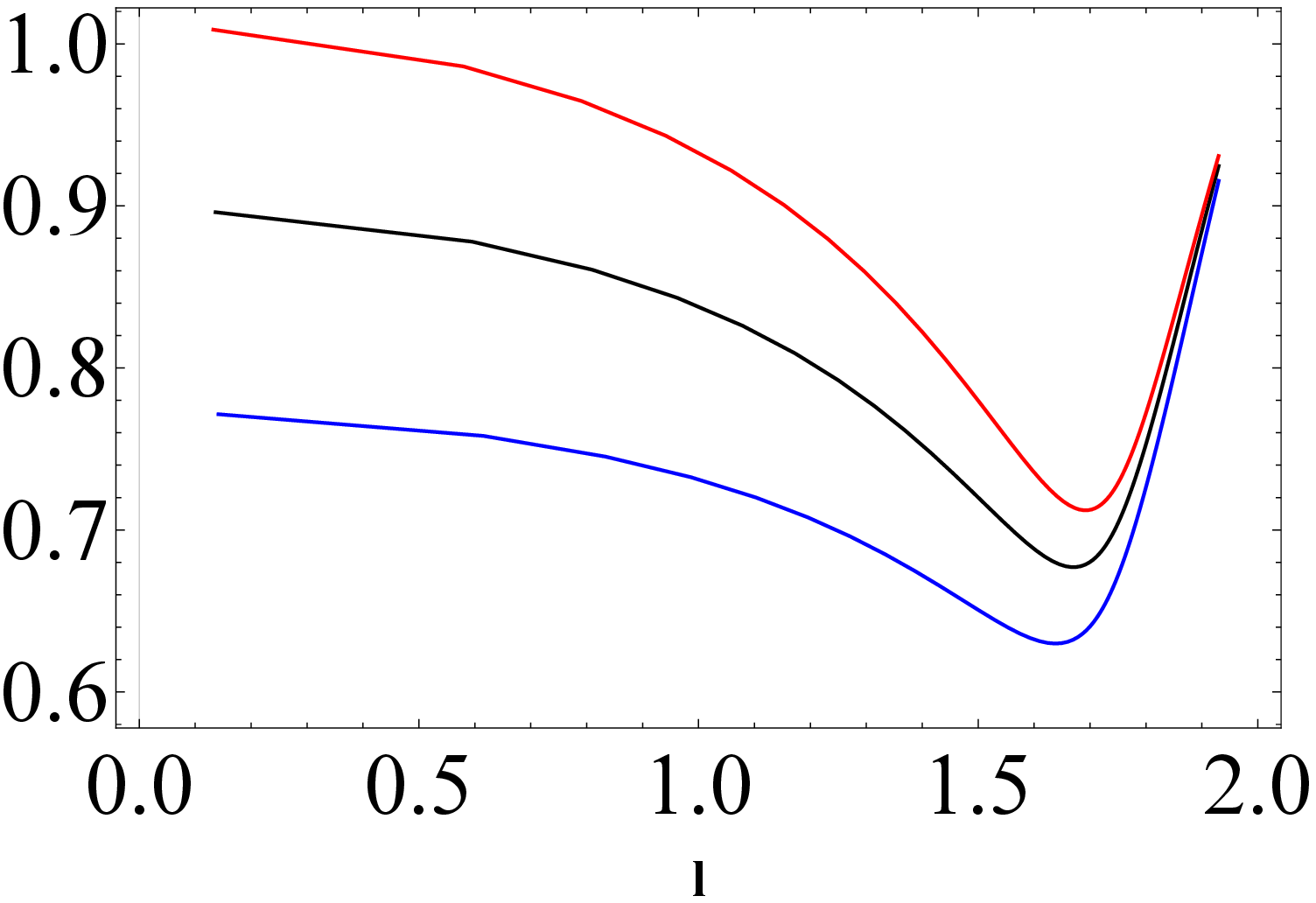}\includegraphics[height=2.5cm,width=5.5cm]{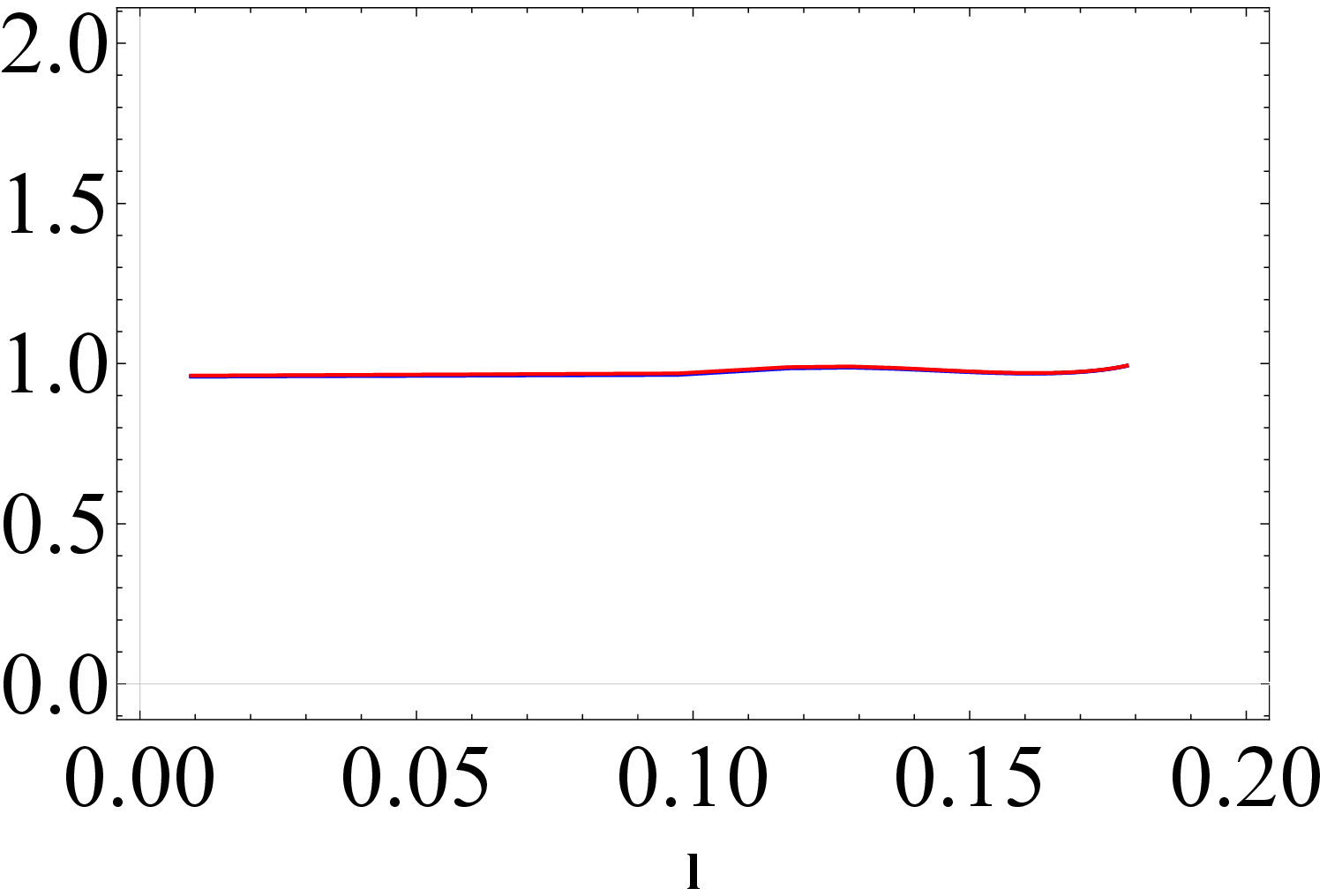}} 
	
	\caption{\label{f4} Top row: energy loss $\Pi$ vs rotation radius $l$; at fixed $l$, energy loss for negative(positive) value of $\lambda_{GB}$ is greater(smaller) than for $\lambda_{GB}=0$. Second row: drag force channel contribution vs $l$ that behaves as energy loss. Third row: Radiation channel contribution vs $l$ that tends to $1$ at large $l$. Bottom row: ratio of $\frac{\Pi}{\Pi _{rad}+\Pi_{drag}}$ vs $l$ that is smaller than $1$ anywhere. In each row, panels from left to right are shown for $\omega=0.05,0.5,5.0$ and at each panel curves from top to bottom are correspond to $\lambda_{GB}=-0.08,0.0,0.08$ respectively.}
\end{figure}

\textbf{\textit{Results:}} We collected all of results on total energy loss and contribution of drag force and radiation channels in Fig. \ref{f4}. In this figure, each row shows a quantity versus rotation radius $l$ where panels from left to right are related to angular velocity $\omega=0.05,0.5,5.0$. Also in each panel, curves from top to bottom are correspond to $\lambda _{GB}=-0.08,0.00,0.08$ respectively. 

In the first row, behavior of energy loss versus rotation radius $l$ is shown. At fixed angular velocity $\omega$ and rotation radius $l$, energy lost by theory with minus (plus) sign of $\lambda _{GB}$ is greater (smaller) than strongly coupled theory associated with $\lambda _{GB}=0$. Here, again we find that at high $\omega$, the gravitational dual theory does not depend on $\lambda _{GB}$ because energy loss curves related to different $\lambda _{GB}$ are matched for $\omega=5.0$. Also in \cite{Fadafan:2012qu} it is shown that, quantities in anisotropic  and isotropic background are the same at high $\omega$. In \cite{AliAkbari:2011ue} authors studied rotating quark in non-conformal background and they found that at the $\omega \to \infty$ limit, there is no difference between low and high temperature  regimes. The reason of it, is understood from Fig. \ref{f2} and the form of $f(u)$ in the background. As it is clear from Fig. \ref{f2}, at $\omega \to \infty$ limit, the worldsheet horizon $u_t$ will be close to the boundary $u=0$, then $f(u_t)\to 1$ in this limit. This leads to the same background in different $\lambda _{GB}$ and high $\omega$.

Second row shows contribution of drag force channel in each $\omega$ and $\lambda _{GB}$. At small angular velocity, almost all of total energy lost by rotating particle is due to drag force and by increasing $\omega$, contribution of drag force is decreasing. At fixed $l$ in each panel, drag force contribution for negative (positive) value of $\lambda _{GB}$ is smaller (greater) than its portion in strongly coupled case. In contrast with the total energy loss, darg force contribution for different backgrounds associated with different $\lambda _{GB}$ are the same at low $\omega$ and will be seperated when angular velocity tends to infinity. It would be reasonable from \eqref{dragcontribution}.

We plotted behavior of ratio \eqref{radiation} in the third row for different $\omega$ and $\lambda _{GB}$. As it is expected, contribution of radiation in total energy loss will be increased by increasing $\omega$ from left panel to right. Also at high $\omega$, energy loss by radiation does not depend on the Gauss$-$Bonnet coupling $\lambda _{GB}$. It is reasonable from \eqref{radiation} where the quantity $\Pi$ and radius rotation $l$ are $\lambda _{GB}-$independent at $\omega \to \infty$ limit.

The ratio $\frac{\Pi}{\Pi_{radiation}+\Pi_{linear drag}}$ is plotted as bottom row. As it is shown, this ratio is lesss than $1$ which means radiation and linear drag force channels to energy loss have destructive  interference between each other. The minimum for two left panels$-$that is a bit greater than $0.6$ $-$ occurs at $l=l_e$ where is solution of $\Pi _{linear~drag}=\Pi_{vacuum~radiation}$ as 

\be \label{minimum}
\omega= \frac{\big(1-l_e^2 \omega ^2 \big) }{\bigg( n (n-l_e^2 \omega ^2)+\lambda _{GB} l_e^4 \omega ^4 \bigg)^{1/4}}.
\ee

Eq. \eqref{minimum} is extracted from \eqref{dragcontribution} and \eqref{rad}, and it is generalized form of the location of minimum come in \cite{Fadafan:2008bq}. The condition  \eqref{minimum} reduces to $\omega = \big(1-l_e^2 \omega ^2 \big)^{3/4}$ for strongly coupled limit $\lambda _{GB}=0$ and $n=1$ that is the same as what is said in \cite{Fadafan:2008bq}. In the regime in which the left hand side of \eqref{minimum} is much greater than the right hand side, the energy loss due to vacuum radiation is larger than linear drag force contribution. But one can say particularly, the radiation contribution will be predominant when $\omega >1$, for each $v$ and each $\lambda _{GB}$, as it is seen in the most right panel in the third row of Fig. \ref{f4}.

\section{Conclusion and Discussion}
We studied a semi classical spiraling string in Gauss$-$Bonnet theory as gravitational  dual to a heavy rotating particle at finite coupling. At the first, we considered string shape in different situation associated with different values of constant of motion $\Pi$, angular velocity $\omega$ and Gauss$-$Bonnet coupling $\lambda _{GB}$. We saw that at fixed $(\Pi,\omega)$, rotation radius is smaller (greater) for negative (positive) values of $\lambda _{GB}$ than strongly coupled case which correspond to $\lambda _{GB}=0$. Based on \cite{Brigante:2008gz}, the ratio viscosity over the entropy density, $\eta / s$ depends on the sign of $\lambda _{GB}$  as $\eta /s=\frac{1}{4\pi}(1-4\lambda _{GB})$. Then when $\lambda _{GB}<0$, this ratio leads to larger values and the theory moves from infinitely strongly coupled regime towards weaker coupling. This is mentioned also in \cite{self2,Grozdanov:2016zjj}. So as it is shown in the second row of figure \ref{f4}, the drag force contribution in total energy loss for negative $\lambda _{GB}$ is smaller than strongly coupled case.\\

\begin{figure}[ht]
	\centerline{\includegraphics[width=10cm]{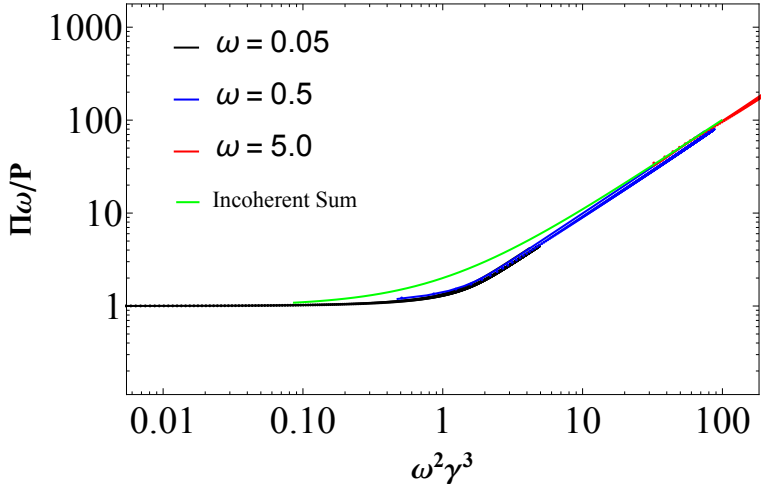} \\ }
	
	\caption{\label{f5} Crossover between regime which predominant contribution of total energy loss comes from drag force channel and regime which the total energy loss is dominated by radiation. The curves associated with minus, zero and plus sign of $\lambda _{GB}$ are matched at all.}
\end{figure}

Also, we found that at  $\omega \to \infty$ limit string shape does not depend on Gauss$-$Bonnet coupling. In addition, considering behavior of worldsheet horizon showed that its radial position in the bulk is $\lambda _{GB}-$independent at high angular velocity and high $\Pi$. Also in limit $\omega \to \infty$, world sheet horizon will be close to the boundary and leads to independence of bulk calculations from $\lambda _{GB}$ because of $f(u_t) \vert _{\omega \to \infty} \to 1$. This independence is expected for calculation  in RN black hole background at high angular velocity. \\

\begin{figure}[ht]
	\centerline{\includegraphics[width=10cm]{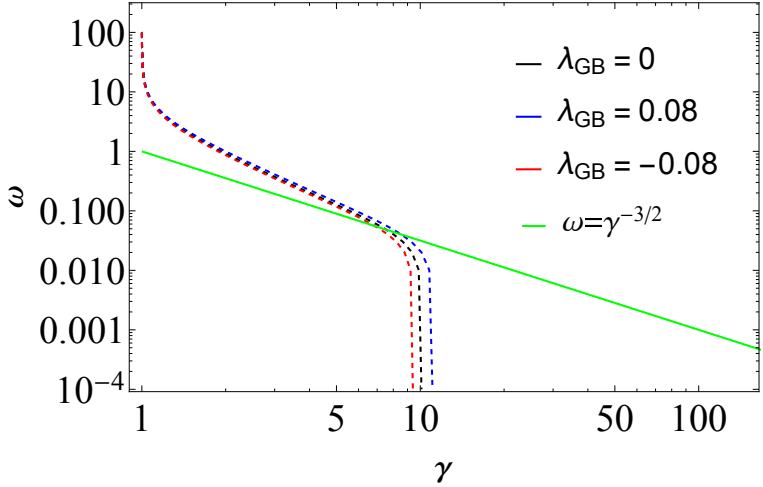} \\} 
	
	\caption{\label{f6} Regime of validity of classical calculation. The dashed red, black and blue curves are correspond to negative, zero and positive values for $\lambda _{GB}$ respectively. This figure is plotted for $\Pi \omega =10$. The solid green line is $\omega =\gamma ^{-3/2}$ that is the same for all values of $\lambda _{GB}$ and shows regime of predominant  contribution of linear drag (below that) and radiation (above that). }
\end{figure}

Next, we computed total energy loss by rotating heavy particle according to its rotation radius. The influence of finite coupling on total energy loss and contribution of drag force and radiation channels is appeared as a shift$-$ depend on sign of $\lambda _{GB}-$ on curves with respect to the strongly coupled case. Here, we saw again $\lambda _{GB}-$independence of calculations at high $\omega$ for total energy loss and radiation  contribution. In contrast, drag force for different $\lambda _{GB}$ were matched  for $\omega \to 0$ limit that was expectable  from \eqref{dragcontribution}.  We also see destructive interference between two channels for different Gauss$-$Bonnet coupling. All of results for $\lambda _{GB}=0$ is in agreement with results of \cite{Fadafan:2008bq}. \\

Figure \ref{f5} shows the crossover between drag and vacuum predominant contribution in total energy loss where energy loss over drag ratio is plotted versus $\omega ^2 \gamma ^3$ with $\gamma$ is given by right hand side of \eqref{dragcontribution}. As it is illustrated in figure \ref{f5}, in regime $\omega ^2 \gamma ^3 \ll 1$ the energy loss is dominated by linear drag limit and for $\omega ^2 \gamma ^3 \gg 1$ the radiation has predominant contribution. This figure is plotted in \cite{Fadafan:2008bq} for energy lost by a heavy rotating particle in $\mathcal{N}=4$ SYM strongly coupled plasma. Here it is repeated for Gauss$-$Bonnet gravity theory with different signs of coupling $\lambda _{GB}$ those are matched  at all. This means that, crossover between linear drag and radiation does not depend on Gauss$-$Bonnet coupling $\lambda _{GB}$.\\

Also, regime of validity is shown in figure \ref{f6}. This is another comparison between finite coupling and strongly coupled cases that shows a simple shift in validity region of classical computation  associated with sign of $\lambda _{GB}$. In this figure, we plotted angular velocity $\omega$ versus relativistic factor $\gamma$ $-$ that is corrected in Gauss$-$Bonnet gravity in R.H.S. of \eqref{dragcontribution}$-$ for $\Pi \omega =10$. The classical calculations  for minus, zero and plus signs of $\lambda _{GB}$ is below the red, black and blue dashed curves in figure  \ref{f6}. The solid green line is $\omega =\gamma ^{-3/2}$ Those are matched for all $\lambda _{GB}$ values. Below this, the total energy loss is due to drag force channel and above that the energy loss comes from radiation. \\

\section*{Acknowledgments}
We would like to thank L. Bianchi, B. Robinson and H. Soltanpanahi for useful discussions and especially thank A. O'Bannon and K. Siampos for the comments.   KBF acknowledges the CERN theory group for its hospitality during the course of this work.


\end{document}